\documentclass{pasj00}
\draft

\newcommand{\magdot}[1]{\ensuremath{^{\rm m}\!\!#1\,}}
\newcommand{\hst}{{\it HST}}
\newcommand{\AAA}{~\ensuremath{\rm \AA \/ }}
\newcommand{\ergl}{~\ensuremath{\rm erg~s^{-1} }}
\newcommand{\ergf}{~\ensuremath{\rm erg~cm^{-2}~s^{-1} }}
\newcommand{\kms}{~\ensuremath{\rm km~s^{-1} }}
\newcommand{\keV}{~\ensuremath{\rm keV}}
\newcommand{\feiii}{{\rm Fe~III}}
\newcommand{\ariv}{{\rm [Ar~IV]}}

\newcommand{\oiii}{{\rm [O~{\sc III}]}}
\newcommand{\sii}{{\rm [S~{\sc II}]}}
\newcommand{\siii}{{\rm S~{\sc III}}}
\renewcommand{\ni}{{\rm [N~{\sc I}]}}
\newcommand{\nii}{{\rm [N~{\sc II}]}}
\newcommand{\niii}{{\rm N~{\sc III}}}
\newcommand{\niv}{{\rm N~{\sc IV}}}
\newcommand{\hei}{{\rm He~{\sc I}}}
\newcommand{\heii}{{\rm He~{\sc II}}}
\newcommand{\pc}{~\ensuremath{\rm pc}}
\newcommand{\Mpc}{~\ensuremath{\rm Mpc}}

\renewcommand{\div}{\ensuremath{-}}
\newcommand{\cmc}{\ensuremath{\rm cm^{-3}\/}}
\newcommand{\Msun}{~\ensuremath{\rm M_\odot}}

\begin{document}
\SetRunningHead{Abolmasov et al.}{ULX Nebula MF16}

\title{Optical Spectroscopy of the ULX-Associated Nebula MF16}

\author{Pavel \textsc{Abolmasov}$^{1,2}$, S. \textsc{Fabrika}$^{1,2}$, O. \textsc{Sholukhova}$^{1}$
\and Taro \textsc{Kotani}$^{3}$}
\affil{$^{1}$Special Astrophysical Observatory, Nizhnij Arkhyz 369167, Russia}
\affil{$^{2}$University of Oulu, P.O. Box 3000, 90014, Finland}
\affil{$^{3}$Dept. of Physics, Tokyo Tech, 2-12-1 O-okayama, Tokyo 152-8551}
\email{pasha@sao.ru, fabrika@sao.ru}       
\KeyWords{ISM: individual (MF16), jets and outflows, supernova remnants
 X-rays: individual (NGC~6946 ULX-1)
 ultraviolet: general
} 

\maketitle

\begin{abstract}
We present the results of optical panoramic and long-slit spectroscopy of the nebula MF16 
associated with the Ultraluminous X-ray Source NGC6946~ULX-1.
More than 20 new emission lines are identified in the spectra.
Using characteristic line ratios we find the electron
density $n_e \sim 600\,\cmc$, electron temperature in the range
from $ \sim 9\,000~\rm K$ to $ \sim 20\,000\,\rm K$ (for different diagnostic
lines) and the total emitting gas mass $M \sim 900 \Msun$.
We also estimate the interstellar extinction towards the nebula as $A_V \simeq
1\magdot{.}54$ somewhat higher than the Galactic absorption. 
Observed line luminosities and ratios appear
to be inconsistent with excitation and ionization by shock waves so we
propose the central object responsible for powering the nebula.
We estimate the parameters of the ionizing source using photon number
estimates and {\it Cloudy } modelling. 
Required EUV luminosity ($\sim 10^{40}$\ergl) is high even if compared
with the X-ray luminosity.
We argue that independently of their physical nature ULXs are likely to
be bright UV and EUV sources. It is shown that the UV 
flux expected in the {\it GALEX} spectral range (1000$\div$3000$\AAA$) is quite reachable for UV photometry. 
Measuring the luminosities and spectral slopes in the UV range may help to distinguish between the 
two most popular ULX models.
\end{abstract}


\section{Introduction}

\subsection{ULXs and their Optical Nebulae}

A point-like X-ray source in an external galaxy
is considered an Ultraluminous X-ray source (ULX) if its luminosity 
exceeds $10^{39} \ergl$ and it is not an active galactic
nucleus. 
It also makes sense to exclude X-ray bright SNe and young 
(such as several years) Supernova Remnants (SNR) that can shine as
bright as $10^{41}\ergl$ in X-rays like the remnant of SN1988Z \citep{FaTer}. 
There are more than 150 ULXs known at the present time \citep{swartz} but very little is
clear yet about the physical nature of these objects.

ULXs are widely accepted as a class of accreting compact objects violating
the Eddington luminosity limit for a conventional stellar mass black hole (about
$1.3 \times 10^{39}$\ergl\ for $10\Msun$).
High luminosity may be a consequence of a higher accretor mass, supercritical accretion, mild geometrical collimation,
relativistic beaming \citep{koer} or some combination of these effects. 
Two models are the most popular:
 {\it (i)} Intermediate-Mass Black Holes (IMBHs) with masses in the range
$10^2\div 10^4 \Msun$ \citep{IMBH,MR2001} accreting at sub-Eddington rates;
 {\it (ii)} supercritical accretion discs like that of SS433 around
 stellar mass black holes observed at low inclinations $i \lesssim 20^\circ$ 
\citep{katz86,king,FaMes}.
Some authors \citep{soria_kuncic} propose ``hybrid'' models involving
$20\div 100~\Msun$ black holes accreting in a moderately supercritical
regime. 
Supercritical accretion discs are considered by different
authors either in the wind-dominated regime first introduced by
\citet{SS73} or in the advection-dominated regime known as slim disc
\citep{ACLS1988,Okajima2007}.

Large number of ULXs are surrounded by large-scale
bubble nebulae \citep{pakull03}, probably shock-powered.
ULX nebulae (ULXNe), however, do not form a homogeneous class of objects.
In \citet{list} we review the observational
properties of 8 ULXNe including the object under consideration. 
We conclude that only about 50\% of ULXNe may be considered
shock-powered shells.
In some of the observed ULXNe high-excitation lines such as \oiii$\lambda$4959,5007 are
enhanced. In some cases HeII$\lambda$4686 emission
is detected as bright as 0.2$\div$0.3 in H$\beta$ units
\citep{lehmann}. 
The line may have stellar origin \citep{kuntz} as well as nebular.
Measuring line widths may help to distiguish
between these two cases.
The nebulae may clarify the nature of corresponding X-ray sources
in two ways:
{\it (i)} probing the photoionizing radiation of the central object
 via gas excitation and ionization conditions;
{\it (ii)} detecting and measuring jet/wind activity by kinematical effects and shock
 emission.

Dynamically disturbed gas may be considered an evidence against the IMBH
hypothesis because a standard accretion disc is unlikely to produce strong 
jets or wind. Supercritical accretion, on the other hand, is supposed
to provide a massive outflow in the form of a strong wind carrying
practically all the accreted mass and having kinematical luminosity 
 $\sim 10^{38}$\ergl\ in the case of SS433 \citep{ss2004}.
Power comparable with the Eddington luminosity ($\sim 10^{39}$\ergl)
may be provided in  the form of relativistic
jets. Their mechanical luminosity acting for $\sim 10^5$ years is sufficient to produce a wind-blown bubble
with properties close to those of some of the existing ULX nebulae \citep{pakull03,list}. 

ULXNe exhibit very diverse observational properties such as
size, morphology and line ratios \citep{list,pakull03}. Integral luminosities
in Balmer lines are in some cases of the order $10^{38}$\ergl\ or higher that
requires an energy source with luminosity $\gtrsim 10^{39}$\ergl\
indicating that ULXs are really powerful objects. 
Therefore strong beaming effects are excluded as the reason of apparently high luminosities of ULXs.
For some objects like HoIX~X-1 and IC342~X-1 the nebula is clearly shock-powered. 
In other cases (HoII~X-1, M101-P098) low intensity of
low-excitation lines, bright \oiii\ and \heii\ emissions and quiet kinematics\footnote{For HoII~X-1 \citet{lehmann}
  report velocity dispersion $\sim 13 \kms$ in the vicinity of the
  X-ray source.} suggest photoionization to
be the main energy source.

The only proven example of a persistent supercritical accretor is the peculiar
binary SS433 \citep{ss2004}. Its accretion rate is of the order
$10^{-4}\Msun\,\rm yr^{-1}$, indicating mass transfer on the
thermal timescale of a massive star. 
Most of the accreting material is ejected in the form of
optically-thick wind with velocities $V \simeq 1000 \div 2000 $\kms.
The total kinetic power of the jets of SS433 is of the order of
$10^{39}$\ergl.
SS433 is surrounded by a large (roughly $40\times 120$\pc) radionebula
W50 \citep{dubner}. The nebula harbors several optical
emission-line filaments situated in the propagation direction of the
jets. 
Large difference in angular size and heavy absorption in the case of
W50 make comparison with ULXNe difficult.

\subsection{NGC6946~ULX-1}

The X-ray source under consideration is known as NGC6946~ULX-1 \citep{swartz},
NGC6946~X-8 \citep{MF16_lira}, N58 \citep{MF16_Holt} or
NGC6946~X-11 \citep{RoCo}.
It was first detected by {\it ROSAT} \citep{schlegel94} 
 and identified with an optical nebula by %
\citet{BF_94} and a marginally resolved radiosource by \citet{vandyk}.  

The nebula is known as MF16 -- Matonick and Fesen~16 \citep{MF_cat}. It was
considered a SNR luminous in X-rays, in fact the
most luminous in X-rays among the optically
bright SNRs \citep{dunne}, till it was proved by \citet{RoCo}
that the X-ray emission originates from a much more compact 
source in the center of the nebula.  {\it Chandra} source coordinates
for J2000 epoch are: $ \alpha = 20^h 35^m 00^s.74$, $\delta = +60^\circ\, 11\arcmin{\,} 30\farcs6$.

The host galaxy NGC6946 is a late-type spiral with active star
formation \citep{degioia_sfr} and the greatest number of detected
supernovae (no less than 8, according to \citet{SN_n6946}). 
The distance to the galaxy is estimated by different authors
as 5.1 \citep{dist_5.1}, 5.5 \citep{dist_5.5}, 5.7 \citep{dist_5.7}
and 5.5$\,\Mpc$ \citep{dist_5.9} therefore we assume $D=5.5\,\Mpc$. 
The divergence in distance estimates leads to a 15\% uncertainty in all
the luminosities inferred.
Spatial scale for $D=5.5\,\Mpc$ is about $27\,\pc$ per arcsecond. 


A point source in the center of MF16 was detected by \citet{BFS}
in {\it HST} data.
The source (star {\bf d} as denoted by \citet{BFS}) has a V magnitude
of $22\magdot{.}64$ and a colour index of $B-V = 0\magdot{.}46$.
The object is significantly redder than
the nearby stars \textbf{a}-\textbf{c} offset to the North-West by several seconds.
If one accounts only for Galactic interstellar extinction (see discussion in
section \ref{sec:charlines}) equal to $A^{(GAL)}_V = 1\magdot{.}14$ according
to \citet{schlegel_abs} in the direction of MF16, the point-like
counterpart resembles an early A~Ia supergiant with $M_V \simeq
-7\magdot{\,}$ and $B-V \simeq 0\magdot{\,}$. 
Additional intrinsic absorption $A_V \sim 0\magdot{.}5$ (reported by \citet{BFS}) places the object
slightly above the Ia sequence in the upper left part of the HR diagram. 

Optical emission-line spectrum of the nebula is generally consistent with
the suggestion of shock heating but also contains high-excitation
lines such as \heii$\lambda4686$ and \oiii$\lambda5007,4959$
doublet \citep{BF_94, BFS}
too intensive for optically-bright SNRs \citep{MFB_opt}.
Partially radiative shocks were proposed to explain the emission-line spectrum
of MF16 \citep{BFS} but they require a very powerful energy
source (see discussion in section \ref{sec:balance}).

The object is much brighter in the optical than
a usual SNR. Its H$\alpha$ luminosity is about $2\times 10^{38}$\ergl, an
order of magnitude higher than the upper limiting H$\alpha$ luminosity for
optically-bright SNRs in nearby galaxies \citep{braunm31}.
\citet{dunne} detected two-component structure in the emission lines of
the object in high-resolution echelle spectra. Broader components
suggest expansion velocities about $250$\kms, narrower components have
velocity dispersion $\sim 20\div 40$\kms.

MF16 is also known as a radio source as bright as $1.3\rm mJy$ at
$20\rm cm$ \citep{vandyk}
marginally resolved by VLA observations as a $\sim 1\arcsec{\,}$ size
object. The radio source appears to be displaced relative to the X-ray
object and the optical source {\bf d} by $\sim 0\farcs5$ (see figure
~\ref{fig:FOV}). Luminosity of the radio
counterpart is about 20 times higher than that of W50 at 20$\rm cm$
\citep{dubner}. The size of the radiosource is poorly known but the
optical nebula is about 5 times more compact than W50. 

MF16 is practically isolated from other HII regions. The nearest one
with comparable brightness is situated about $200\pc$ away \citep{BFS}.
The inner brighter part of the nebula has a shape of a shell $20 \pc
\times 34 \pc$ elongated in East to West direction with a brighter
Western loop. In the deep {\it HST} images a faint asymmetric halo may be seen
\citep{BFS}. 

In the following section we describe the observational data. In
section~\ref{sec:spectra} we present the results of spectral analysis.
In section~\ref{sec:balance} we analyze the
excitation and ionization sources of the nebula and model the observed spectrum using
{\it Cloudy}. In section~\ref{sec:trep} we discuss our results and the
perspectives of UV observations of ULXs.

\section{Observations and Data Reduction}

All the data were obtained at the Russian SAO 6m telescope.
The main information about the observational material is listed in
table~\ref{tab:obstab}.

The data were flat-fielded and wavelength-corrected. Wavelength
calibration was made using a He-Ne-Ar lamp. 
Atmospheric extinction 
was corrected using
the data obtained by \citet{ext78} fitted by a simple
formula $\Delta m = (0.013\,(\lambda/10^4\AAA)^{-4}+0.13) \sec\, z$.

\subsection{Integral-Field Spectroscopy}

The MultiPupil Fiber Spectrograph (MPFS) is described in \citet{MPFSdesc}.
MPFS is a panoramic spectrograph with
$16\arcsec{\,}\times 16\arcsec{\,}$ field of view consisting of
256 $1\arcsec{\,}\times 1\arcsec{\,}$ spatial sampling
elements. Spectra from individual elements arranged according to their
coverage of the celestial sphere form a three-dimensional structure
that is usually called a data cube. 
Extra 17 fibers collect sky background from $4\arcmin{\,}$ away offset regions.


Data reduction system was developed in IDL environment. The reduction process    
includes the procedures standard for panoramic spectroscopy data
reduction (see for example \citet{pmas_sanchez}): bias subtraction, cosmic hits removal, flat-fielding,
individual fiber extraction, wavelength and flux calibrations.
The dispersion curve has an accuracy of about  
$5-10$\kms. Wavelength shifts between different fibers were corrected
using the brightest night sky lines [OI]$\lambda\lambda$5577,6300,6364.
We used GD248 spectrophotometric standard \citep{oke90} for
flux calibration. Sensitivity variations of individual fibers were
corrected using twilight sky exposures. 

We also added atmospheric dispersion (or atmospheric differential  
refraction, ADR) correction with accuracy better than              
$0\farcs1$ in order to achieve correct relative astrometry for the 
datacube. We determined the value of ADR using the spectral standard
star. Its shift along the vertical as a function of wavelength 
was approximated by a second-order polynomial that resulted in a
roughly $0\farcs1$ scatter around the best-fit parabola and no
detectable systematical deviations.
The spectral standard was observed at different zenith distance
($z_{star} \simeq 40^\circ$, $z_{obj} \simeq 55^\circ$) but we corrected
for that difference using the well-checked  $\propto \tan z$
dependence for ADR. 
According to \citet{filippenko}  $\propto \tan z$
approximation holds within $\sim 0.2\%$ at the $2~km$ altitude in the
optical range up to $\sec z \sim 5$ ($z \sim 80\deg$).




\subsection{Long-Slit Spectroscopy}

\begin{figure*}
\centering
\FigureFile(15cm,8cm){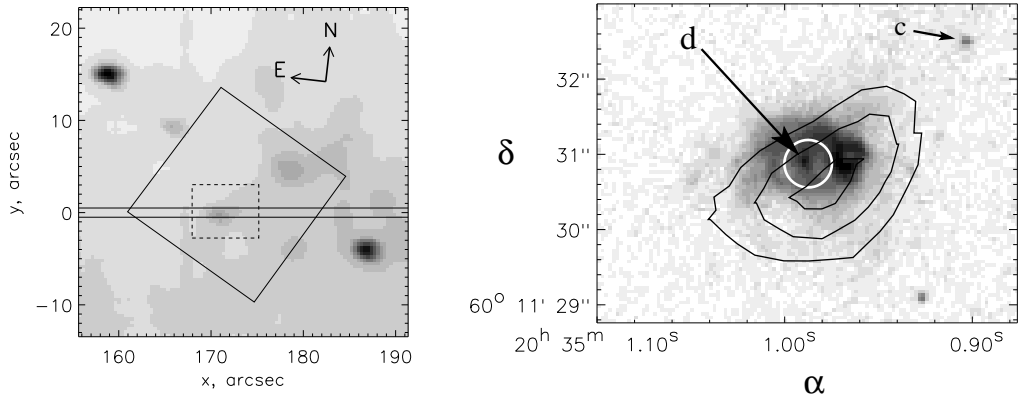}  
\caption{V-band image obtained during the observations
with SCORPIO (on the left panel) and HST ACS archival image (F658N filter) of the
source vicinity (right panel, the region is shown on the left panel by
a dashed rectangle). MPFS field is shown by a square. 
Slit position is shown on the left panel by two horisontal
lines. Point sources {\bf c} and {\bf d} 
are marked. White circle is centered at the position of the X-ray source,
black contours correspond to the VLA radio isophotes at 4.86GHz \citep{vandyk}.
}
\label{fig:FOV}
\end{figure*}

The observations were held with SCORPIO focal reducer \citep{scorpio}, 
in the long-slit mode with moderate spectral resolution
and slit width $1\arcsec{\,}$. 
The position angle ($PA = 96^\circ\!.5$) of the slit was chosen
along the major axis of the nebula (figure~\ref{fig:FOV}).
Reduction was performed in IDL environment and includes all the
standard reduction steps for long-slit spectroscopy. 

The signal-to-noise 
ratio for the long-slit data is about factor 2-3 higher than in the 
MPFS data, mainly because of the better spectrograph transparency. 
Integral field data, however, have the advantage of giving flux
values free from slit losses and radial velocities without
systematical errors usual for long-slit spectroscopy.
Besides this, MPFS data cover larger spectral range.

\section{Spectroscopy Results}\label{sec:spectra}

\subsection{Emission Line Spectrum}\label{sec:lines}

MPFS spectrum was extracted within $2\arcsec{\,}$ radius diaphragm,
the long-slit spectrum was integrated within a $\pm 1\farcs5$ band.
Emission line fluxes were calculated using gauss analysis. Fitting with
two gaussians was used for [SII]$\lambda\lambda$6717,6731 doublet,
H$\gamma$+[OIII]$\lambda$4363, HeII$\lambda$4686+FeIII$\lambda$4658
and FeII+FeIII~$\lambda$5262,5270 blends and line groups.
Triple gaussian was used to deblend H$\alpha$ with
[NII]$\lambda$6548,6583 doublet (for the components of the
latter fixed wavelength difference and flux ratio 
$F(\lambda 6583 / \lambda 6548)=3$ were adopted). 
Exact (accuracy $\sim 0.1\AAA$ or better) wavelengths were taken from
\citet{coluzzi}.

Table~\ref{tab:lines} lists the parameters of all the detected
emission lines in the extracted spectra.
All the lines redward of 5700\AAA\ are from the MPFS data,
most of the bluer lines have fluxes measured with SCORPIO.
Radial velocities from the panoramic data are given when possible. 
Number of lines such as \ni$\lambda$5200 and \sii$\lambda$4068 are close doublets or blends
with unknown component ratios and therefore may have biased radial
velocities by several tens of \kms.
Errors in table~\ref{tab:lines} give the uncertainties of gaussian
parameters. Signal-to-noise ratio is $S/N \sim 3$ for the faintest
lines identified like \feiii$\lambda$4936.
All the fluxes in the second column are normalized over the integral H$\beta$
flux in the corresponding spectrum, and unreddened fluxes over the
unreddened H$\beta$ flux value.
Total flux from the MPFS data is $F(H\beta) = (4.66\pm 0.13)\times
10^{-15} \ergf$. Unreddened values are calculated using 
the reddening curve by \citet{CCM} with $R_V = 3.1$ and the
interstellar extinction value
$A_V = 1\magdot{.}54$ calculated in section ~\ref{sec:charlines}. 
Unreddened H$\beta$ flux is $F_{(unreddened)}(H\beta) =
(2.50\pm0.07)\times 10^{-14} \ergf$.

Single strong lines show non-gaussian profiles possibly broadened by
kinematical effects of the order $200\kms$. We however lack spectral
resolution to deblend the broad and narrow components detected by
\citet{dunne}.
Some of the line identifications are debatable, mostly for weaker lines.
However, we definitely see a number of high-excitation emissions like
Pickering \heii$\lambda$5412 and two putative \ariv\ lines.
Lines of moderately high excitation like \hei, \feiii\ and \siii\ dominate
in number among the weaker lines. 
Some of the emissions are forbidden and therefore of nebular origin.
FWHM of the \heii$\lambda$4686 emission does not differ significantly
from these of the brightest nebular lines like \oiii$\lambda$4959,5007
hence \heii$\lambda$4686 is likely to be a nebular emission as well. 
Stellar or accretion disc emission 
is expected to be broadened by about $1000\kms$ \citep{wr_hamann}.
Wolf-Rayet \niii-\niv\ and CIII/CIV blends at about 4640\AAA\ are definitely absent
(at least fainter than the \feiii$\lambda$4658 emission) also
supporting nebular origin of the \heii$\lambda$4686 emission.

\begin{figure*}
 \centering
\FigureFile(160mm,200mm){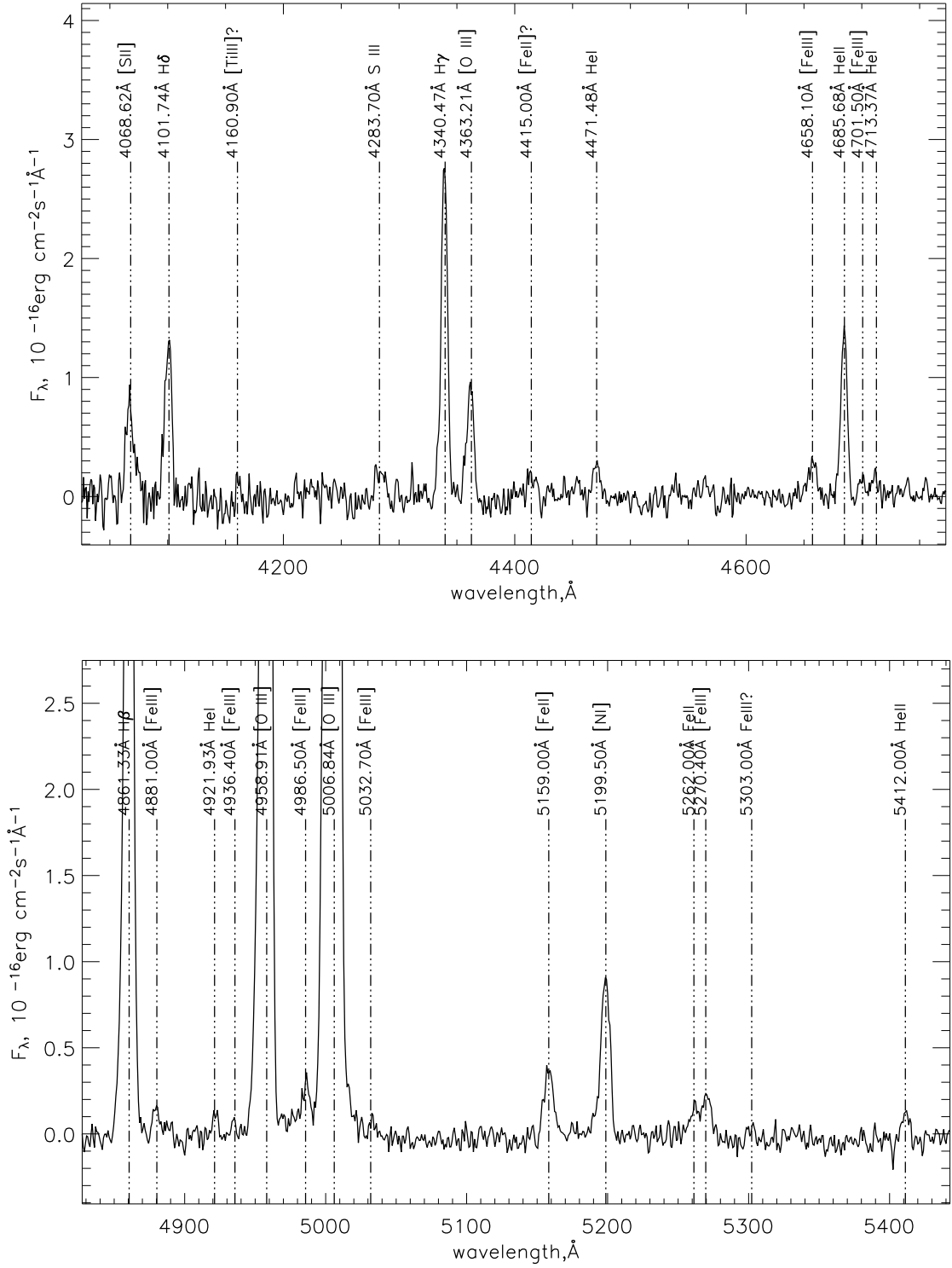}  
\caption{MF16 integral spectra extracted from the SCORPIO
long-slit data.}
\label{fig:spectra}
\end{figure*}

\subsection{Characteristic Line Ratios}\label{sec:charlines}


H$\alpha$/H$\beta$ flux ratio is commonly used as a probe for interstellar absorption
because of its very little dependence on all the plasma parameters.
The ratio is nearly constant and equal to 3 with $\sim$10\% accuracy in the
$5\,000\div 20\,000\,\rm K$
temperature and $1\div 10^{4}\,\cmc$ number density range for photoionized plasma.
We used \citet{CCM} extinction curve with $R=3.1$.
The integral MPFS spectrum gives $A_V = 1\magdot{.}34$ that is slightly
greater than the Galactic absorption $A^{Gal}_V = 1\magdot{.}14$ in
the X-ray source direction \citep{schlegel_abs}. 
H$\alpha$/H$\beta$ ratio might be higher than 3 \citep{DoSutISM} in case of collisional
excitation. The effect may be important, for example, for shock waves
of moderate velocities $V_S \lesssim 100$\kms\citep{shullmckee}, but
its impact is questionable and virtually unstudied. 
Using the electron temperature and density estimates calculated below this
section ($T \simeq 15\,000\,\rm K$, $n_e\simeq 600\,\cmc$) one arrives to a slightly lower
H$\alpha$/H$\beta$ ratio $\sim 2.8$ that corresponds to $A_V \simeq
1\magdot{.}54$ rather than the value stated above.


Previous spectral studies of MF16 by \citet{BF_94} and \citet{BFS}
report higher values of interstellar absorption: $E(B-V)=0.52$
(implying $A_V\simeq 1\magdot{.}6$) and  $E(B-V)=0.65$
($A_V\simeq 1\magdot{.}8$), correspondingly. 
Possible explanation is in  patchy extinction making the $H\alpha /
H\beta$ flux ratio variable within the extent of the nebula.
This is consistent with the relatively low intrinsic absorption
(intrinsic colour excess $E(B-V) \lesssim 0\magdot{.}1$, implying $A_V
\lesssim 0\magdot{.}3$) estimated for nearby
stars {\bf a}-{\bf c} \citep{BFS} and high absorbing column obtained for the
X-ray source by \citet{RoCo} $N_H \sim 4 \times
10^{21} cm^{-2}$ corresponding to $A_V \sim 1\magdot{.}8$
\citep{gorenstein}.



In order to probe the physical conditions in the nebula we used
several characteristic emission line ratios. Characteristic
line calibrations were made according to \citet{osterbrock} and
{\it STSDAS} on-line {\it NEBULAR.temden} resource at {\it http://stsdas.stsci.edu/nebular/temden.html}.

\sii\ 6717 / 6731 intensity ratio is usually used for electron
density estimates.
Integral line flux ratio ($1.05\pm 0.03$) indicates a rather high value, $n_e =
570\pm 60 \cmc$ if one assumes $T=15\,000\rm\,K$ (see below). 
The electron temperature is usually estimated using two characteristic line ratios:
$ \oiii\, \left( \lambda 4959 + \lambda 5007 \right) / \lambda 4363$
and $\nii \left( \lambda 6548 + \lambda 6583 \right) / \lambda 5755$.
For different ions we estimate electron temperature as: $T(\oiii) = 17\,700\pm1200\,\rm K$,
$T(\nii) = 15\,600\pm2\,000\,\rm K$. Somewhat hotter temperatures inferred
from \oiii\ lines are quite reasonable, because \oiii\ has higher ionization
potential and emits in regions with higher temperature.
An estimate for the coldest regions of the nebula may be made using
$\sii (\lambda 4068+\lambda 4076) / (\lambda 6717+\lambda 6731)$ flux ratio. Using {\it
  temden} we arrive to a temperature estimate $T(\sii) = 8\,600\pm
500\,\rm K$. Large wavelength difference makes this value less defined due
to uncertainty in the interstellar absorption value, $T(\sii) =
9\,000\pm 1\,000\,\rm K$. Here, absorption uncertainty $\sim 0\magdot{.}2$
results in about $12\%$ flux ratio uncertainty for the
temperature-sensitive doublets of \sii\ making total flux ratio uncertainty about
two times higher. Accurate temperature determination is limited by the
variable absorption effects mentioned above. 

Total luminosity of the nebula in the H$\beta$ emission line may
be used to estimate the emission measure and the total mass of the
emitting gas. 
 If one assumes constant density and temperature of the order $(1\div 2) \times
10^4\,\rm K$ for all the ionized material the mass
of the nebula may be expressed as follows:

\begin{equation}\label{E:hbmass}
\begin{array}{l}
  M \simeq \frac{m \, EM}{n_e E(H\beta)} \simeq \frac{m L(H\beta)}{\alpha^{eff}(H\beta) n_e E(H\beta)},
\end{array}
\end{equation}
\noindent

where $m$ is the mean particle mass and $\alpha^{eff}$ is the effective (Case B) recombination
coefficient. $L(H\beta) = (9.0\pm 0.2) \times 10^{37}\ergl$ is the object
luminosity is H$\beta$.
Using the estimates for $n_e = 600\rm\, cm^{-2}$ and $T_e = 15\,000\,\rm K$
one yields for the case of MF16  $M = 870 \pm 50 \Msun$, about two
times higher than the value obtained by \citet{BFS} because of higher
H$\beta$ luminosity inferred. That is similar to the mass of the
shovelled material for a spherical nebula with $R = 13\,\pc$\ and
original interstellar hydrogen density $n_0 \simeq 6 \cmc$.

\section{Ionization sources}\label{sec:balance}

\subsection{Shock Excitation}

Most of the observational properties of MF16 such as two-component emission
 lines \citep{dunne}, morphology \citep{BFS} and high
 \sii$\lambda\lambda$6717,6731 / H$\alpha$
 flux ratio favour shock excitation.
 \citet{DoSutI} provide a useful formula for the total flux emerging
 from a unit radiative shock front surface area in 
H$\beta$ (both for the post-shock flow and for precursor):

\begin{equation}\label{E:hbeta}
 \begin{array}{l} 
F_{H\beta} = F_{H\beta,shock}+F_{H\beta,precursor} = \\ 
\qquad{}
\left( 7.44 \times 10^{-6} \left(\frac{V}{100\kms}\right)^{2.41}+\right.\\
\qquad{}
\left. + 9.86 \times 10^{-6}
\left(\frac{V}{100\kms}\right)^{2.28}\right) \times \\
\qquad{}
\times \frac{n_0}{1\,\cmc} \ergf \\
 \end{array}
\end{equation}
\noindent

Here $n_0$ is the pre-shock hydrogen density
and $V$ is the shock velocity in $100$\kms\ units.
The estimates are supposed to be valid for shock velocities of the
order $100-1000$\kms\ and densities usual for the ISM and nebulae
($0.1 \div 10^4 \cmc$). Precursor emission (the second term in brackets) 
is present only for shock velocities $V_S \gtrsim 150$\kms\ because
slower shocks are unable to produce an 
ionization front moving fast enough \citep{DoSutISM}.

Integrating (\ref{E:hbeta}) over shock fronts gives total H$\beta$ luminosities
that may be compared with the observed luminosity of the emission-line object.
The integral unreddened H$\beta$ luminosity following from the 3D data analysis
(see section~\ref{sec:lines}) is
$L(H\beta) = (9.0\pm0.2)\times10^{37}\,\ergl$ or
$(2.50\pm0.07)\times 10^{-14}$\ergf\ in flux units. 
Formula ~(\ref{E:hbeta}) may be rewritten as:

\begin{equation}\label{E:velpow}
\begin{array}{l}
F(H\beta) = 
\left( 2.0 \left(\frac{V}{250\kms}\right)^{2.41} + \right. \\
\qquad{} \left. + 2.4 \left(\frac{V}{250\kms}\right)^{2.28} \right)\
\frac{n_0}{5\,\cmc} \ \theta^2 \times \\
\qquad{} \times 10^{-15} \ergf
\end{array}
\end{equation}
\noindent
$\theta$ is the angular diameter of the nebula in arcseconds.  
The estimate (\ref{E:velpow}) predicts H$\beta$ flux several times less than the observed value.
Shock waves are short in producing enough flux in H$\beta$ unless 
the initial density $n_0 > 10~\cmc$ (that is unlikely because the
mass of the shocked material should be much higher than the mass of
the emitting gas) or the shock velocity $\gtrsim 300$\kms\ (that is
excluded by kinematical data). 
 Besides this, observed intensities of \oiii\ and \nii\ lines require shock with $V_S \sim
300\div 400\kms$ \citep{DoSutI,evans_shocks} that is significantly
higher than what one may allow having the kinematical limits
mentioned above. 
\heii$\lambda4686$ emission line is much  brighter with respect to H$\beta$ than
a shock wave with precursor with $V_S < 1000 \kms$ can provide 
\citep{evans_shocks}. 
We conclude that observed expansion velocities are
too low to explain the emission line luminosities observed as well as
the line ratios. 

\subsection{Photoionization}

An evident solution is to consider a bright photoionizing source that
can increase the total luminosity in H$\beta$ without changing the line widths. 
One may suggest the X-ray radiation of the central source responsible
for powering the nebula. In order to check the ability of the X-ray
source to power the nebula we set the radius of the nebula equal
to 13pc (according to \citet{BFS}) and used the best-fit MCD + power-law model from \citet{RoCo}. 
MCD (MultiColour Disc blackbody) was
calculated according to \citet{discbb}. 
Soft MCD component was extrapolated to the EUV region, but we 
truncated the power-law at $0.5~\keV$ where its intensity becomes comparable with that of the
MCD component. 

The observed X-ray radiation provides an ionization parameter
logarithm $\lg U \simeq -5.2$
\footnote{Here and below we define ionization parameter as (following \cite{evans_shocks}):
$$ U = \frac{1}{cn_H} \int^{+\infty}_{13.6\rm eV\/ }\frac{F_{\nu}}{h\nu} d(h\nu),$$
\noindent
where $n_H$ is the total (ionized + neutral + molecular) hydrogen
density in the gas.}.
However, observed \oiii$\lambda$5007/H$\beta \sim 7$ and
HeII$\lambda$4686/H$\beta \sim 0.2$ ratios point to a rather high
ionization parameter logarithm value $\lg U \gtrsim -3$ \citep{evans_shocks}.
Besides this, as we see below, the X-ray source itself is unable to produce enough photoionizing quanta.
This can be done by introducing a powerful extreme ultraviolet (EUV)
source ($L_{UV} \sim 10^{40} \ergl$).


\heii$\lambda4686$ emission line is much  brighter with respect to H$\beta$ than
a shock wave with reasonable parameters may explain.
Total luminosity in the line is $L_{\heii\lambda 4686} = (2.0 \pm 0.2)
\times 10^{37}\ergl$ according to the MPFS data.
Because of this and the high excitation potential of the corresponding 
transition, we consider \heii$\lambda$4686 a recombination line.
One can estimate the number of ionizing quanta required and the luminosity of the photoionizing source following
\citet{osterbrock} (Zanstra method):

\begin{equation}\label{E:heion}
\begin{array}{l}
Q(He^+) \ge  \frac{1}{E(\lambda\,4686)} \times \\
\times \frac{\alpha_B}{\alpha^{eff}(HeII\,\lambda4686)} \ \  L(HeII\lambda\,4686), \\
\end{array}
\end{equation}

\noindent
$\alpha_B$ and $\alpha^{eff}(HeII\,\lambda4686)$ here are the
total recombination rate for the Case B and the effective recombination rate
for the HeII$\lambda$4686 line \citep{osterbrock}, correspondingly. 
Their ratio depends weakly on plasma parameters,
varying by a factor of 2 in the plasma temperature range $(1\div30) \times 10^3\,\rm K$. 
For particle density of the order $100-1000\,\cmc$ and temperature
$T \sim (1.5\div 2)\times 10^4\,\rm K$ 
the recombination coefficient ratio
 $\alpha_B / \alpha^{eff}(HeII\,\lambda4686) \simeq 4.7$.
 Corresponding luminosity of the He$^+$-ionizing EUV source is 
$L(\lambda < 228\AAA) \gtrsim 95 L(HeII\lambda\,4686) \simeq 2 \times 10^{39}$\ergl. 
Note that all the ionizing luminosity must be concentrated in the wavelength range $20-200\AAA$
(at $\lambda \lesssim 20\AAA$ heavier elements dominate the
extinction). The luminosity is calculated for $\lambda=228\AAA$ and
actually is a lower estimate because of higher photon energies and
quanta leakage. 
Though bright HeII$\lambda$4686 emission seems to be quite usual for
ULX counterparts \citep{list}, for MF16 the HeII luminosity is
extremely large. 
The EUV luminosity inferred is comparable with the X-ray luminosity.

Similar estimates may be made for other recombination lines such as
H$\beta$ and HeI$\lambda$4471.
%
The Case~B numbers of $He^+$, $He^0$ and $H^0$-ionizing quanta derived from
HeII$\lambda$4686, HeI$\lambda$4471 and H$\beta$, correspondingly, are
as follows:

\begin{equation}\label{E:heiiQ}
Q(He^+) =  (2.2\pm 0.2)\times 10^{49}
\frac{L(HeII\lambda\,4686)}{2.0 \times 10^{37} \ergl} \ s^{-1}
\end{equation}
\begin{equation}\label{E:heiQ}
Q(He^0) =  (1.8\pm 0.2)\times 10^{49}
\frac{L(HeI\lambda\,4471)}{4.0\times 10^{36} \ergl} \ s^{-1}
\end{equation}
\begin{equation}\label{E:hydQ}
Q(H^0) =  (1.90\pm 0.05)\times 10^{50}
\frac{L(H\beta)}{9.0\times 10^{37} \ergl} \ s^{-1}
\end{equation}

\noindent
The ionizing quanta numbers listed above correspond to ionizing source
luminosities greater than $10^{39}$\ergl. Luminosity required to
explain the $H\beta$ luminosity of MF16 is $L(\lambda < 912\AAA)
\gtrsim 4 \times 10^{39}\ergl$.

\subsection{Photoionization grid}\label{sec:grid}

We have computed a grid of {\it Cloudy} version 07.02.00 \citep{cloudy98} photoionization models
in order to fit the spectrum of MF16 neglecting shock waves. We assumed
all the plasma situated at $13\,\pc$ from the point source in the center,
forming a hollow envelope with hydrogen density $n_H = 500\,\cmc$.
We considered photoionization with an isotropic source
with a two-component spectrum: fixed X-ray component observed by {\it
Chandra} (and extrapolated in the EUV region in the same way it was
done in the previous section)
and a blackbody source with the temperature spanning the range from $10^3$ to $10^6\,\rm K$ and 
integral flux densities from 0.01 to 100\ergf\
(corresponding to the luminosity range $10^{37}\div 10^{41}$\ergl). 

8 lines with different ionization potentials were used for fitting. 
Measured emission line fluxes were fitted assuming the distance
$D=5.5\,\pc$. 
Two interstellar extinction values were applied, $A_V = 1\magdot{.}34$
and $1\magdot{.}54$, the latter resulting in much better fit
supporting the idea that the flux ratio H$\alpha$ / H$\beta\, \simeq 2.8$ rather than
$3$. Fitting results and partial $\chi^2$ values are given in
table~\ref{tab:linefit}. 



The best-fit parameters for MF16 are: $\lg T(K) = 5.15 \pm 0.10$, $F = 0.6 \pm
0.1\,\ergf$ resulting in a normalized $\chi^2 \simeq 25$ (for 6
degrees of freedom). The luminosity of the corresponding 
UV source is $L_{UV} \simeq 1.2 \times 10^{40} \ergl$
(mainly concentrated at $200-400\AAA$).
The high values of the reduced
$\chi^2$ are a direct result of the high S/N in our data.
 Note that all the line fluxes are predicted
with the accuracy better than $\sim 20\% $ quite reasonable for our
analysis. All the line ratios are affected by
plasma inhomogeneities, abundance
and depletion effects. In principle a more complicated
model may be applied to our data.
Hotter lines like \heii$\lambda$5412 and \oiii$\lambda$4363 are
enhanced in the observed spectrum with respect to the best-fit photoionization
model predictions indicating an additional source of heating. Model \oiii\
temperature is about $10^4\,\rm K$ instead of $1.7\times 10^4\,\rm K$ measured
from the  \oiii\ characteristic line ratio. 

Our models also overestimate the intensities of \feiii\ lines.
This may be attributed to partial depletion of the
element into dust taking place in the interstellar medium.
Iron depletion values are usually of the order $10^{-2}$ for
undisturbed dense ISM and about $0.1$ for HII-regions
\citep{rodriguez2002} indicating that dust is effectively destroyed
in MF16. 
Shock waves are likely to destroy significant part of dust grains
in the ISM but some percentage of grains survive for $V_S \lesssim
500\kms$ \citep{DoSutISM,Jones94}. The rich spectrum of \feiii\
indicates that the emitting gas was processed by shock waves showing
however depletion by a factor of $0.3\div 0.5$. 


We conclude that the observed emission-line spectrum may be reproduced without any
significant abundance changes and without shock waves. 
However, the
effect of the latter may still be important (in heating
the medium and destroying dust). 



\section{Discussion}\label{sec:trep}

\subsection{The Nature of the EUV Source}\label{sec:euvsrc}

Cool multicolour disc (MCD) + power law models are often used to fit
ULX spectra and are considered an argument for the IMBH model \citep{IMBH}.
Hard power law component usually dominates the X-ray spectrum but it is expected to be
truncated at energies close to the inner disc temperature. For
NGC6946~ULX-1 about 20\% of the standard {\it Chandra} X-rays may be attributed to the
thermal MCD component with the inner temperature $T_{in} = 0.15\,\keV$ \citep{RoCo}. 
Bolometric luminosity of 
a standard disc with such a temperature should be about 5 times higher than
its luminosity in the standard {\it Chandra} X-ray range. 
Though most of the disc luminosity is emitted in the EUV range
the number of photons is insufficient to explain the Balmer lines.
It may be checked that a conventional MCD with the parameters measured
by {\it Chandra} provides only $Q_H \simeq 10^{49}\,\rm s^{-1}$ -- at least an
order of magnitude less than is needed to explain the H$\beta$
luminosity. Luminosity of the \heii$\lambda$4686 line is however only
about 2 times higher than the MCD component can provide. 


The central optical point source (star \textbf{d} in figure~\ref{fig:FOV}) is
much brighter than a possible low-energy tail of a
standard accretion disc. According to \citet{BFS}, star \textbf{d} has
$V=22\magdot{.}64$ and $B-V = 0\magdot{.}46$.
For a distance of $5.5Mpc$ and $A_V = 1\magdot{.}54$ the central optical point-like source will have
$M_V = -7\magdot{.}6$ ($L_V \sim 3 \times 10^{38}$\ergl) instead of $M_V = -2\magdot{.}7$ given by the best-fit
MCD for {\it Chandra} data with an infinite outer radius. The optical
excess, however, may be attributed to the donor star.
The optical object itself is relatively cool: $(B-V)_0 \simeq
-0\magdot{.}05$ if one assumes colour excess following from the
$A_V$ estimates due to H$\alpha$/H$\beta$ ratio $E(B-V) \simeq A_V /
3.1 \simeq 0.5$. That corresponds to a temperature $\sim
10^4K$. Exponentially small part ($\exp\{-Ry / T\} \sim 4 \times 10^{-5}$) of the photons emitted by the
star is able to ionize hydrogen and helium, therefore the contribution
of the donor star to the ionization balance in MF16 is probably
negligible.  
About $ 1\magdot{\,}$ additional intrinsic absorption
is needed to make the optical source a hot blackbody with $B-V \sim
-0.3 \magdot{\,}$. 

In figure~\ref{fig:sedgalex} we reconstruct the spectral energy
distribution (SED) of the ULX throughout the UV spectral range (from X-rays to the optical). 
Our estimates of 
ionizing luminosities put together with the X-ray best-fit
model spectrum \citep{RoCo} result in
a source with an approximately flat ($\nu L_\nu \sim const$) spectrum
from 0.01 to 10$~\keV$. 
Ionizing flux estimates (\ref{E:heiiQ}-\ref{E:hydQ}) are indeed
lower limits due to higher energies of the absorbed quanta and possible quanta leakage.
The estimated EUV luminosity can not be explained by an MCD with $T_{in} \sim
0.1\div 0.2\,\keV$.
In figure~\ref{fig:sedgalex} we show MCD spectra
for different black hole masses (10, 100, 1000 and 10000$\Msun$
from bottom left to top right) accreting at 1\% of the critical
accretion rate. 

For a standard disc \citep{SS73} luminosity and temperature scale with the black
hole mass $M$ and dimensionless accretion rate $\dot{m}$
as $L \propto M \dot{m}$, $T_{in} \propto M^{-1/4} \dot{m}^{1/4}$, correspondingly. Mass
and accretion rate estimates are therefore much more
sensitive to the SED shape than to luminosity: $M \propto L^{1/2}
T_{in}^{-2}$, $  \dot{m} \propto  L^{1/2} T_{in}^2 $. The black hole mass in
the framework of the IMBH hypothesis should be $\sim (1\div 3) \times
10^4 \, \rm M_{\odot}$. 

Our best-fit model parameters (section~\ref{sec:grid}) suggest an
object with radius $\sim 10\,\rm R_\odot$ and temperature $\sim
10^5\,\rm K$
too high even for a conventional Wolf-Rayet star. Hot WC and WO stars
may have comparable effective temperatures, but their luminosities and
ionizing quanta production rates are at least an order of magnitude
lower \citep{WR_crowther}. 
A possible interpretation is emission from the optically thick wind
of a supercritically accreting stellar mass black hole that should be a bright UV/EUV source
of appropriate luminosity \citep{poutanen}.
The X-ray component may be attributed to the funnel radiation in that case.
\citet{poutanen} estimate the outer photosphere temperature of a
supercritical accretion disc wind as:

\begin{equation}\label{E:windph}
T_{ph} \simeq 0.8 m^{-1/4} \dot{m}^{-3/4} \,\keV
\end{equation}

\noindent
where $m$ is the black hole mass in solar units, $\dot{m}$ is
the dimensionless mass accretion rate at the infinity. From
the equation (\ref{E:windph}) the dimensionless mass accretion rate may be estimated
as $\dot{m} \simeq 200 \, m_1^{-1/3} T_5^{-4/3}$, where $m_1$ is the black hole mass in $10\,\Msun$\
units, and $T_5$ is the temperature of the photosphere of the wind in $10^5\,\rm K$.
The accretion (or, more strictly, mass ejection) rate is found to be about an order of magnitude
less than expected for SS433 \citep{ss2004} yet still highly supercritical.

\subsection{ULXs in the UV}

\begin{figure*}
\centering
\FigureFile(160mm,80mm){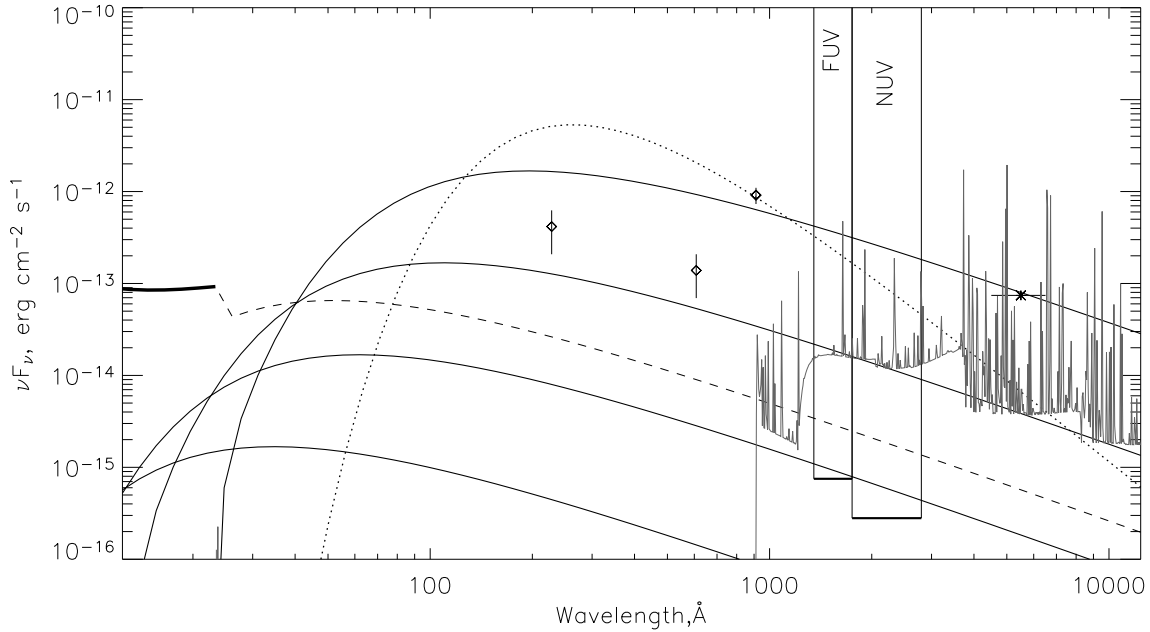}
\caption{ 
Reconstructed SED of NGC6946~ULX-1 from X-rays to the optical.
  Part of the best-fit X-ray spectrum \citep{RoCo} is shown by a thick solid line at shorter
  wavelegths. Dashed curve shows the behavior of the {\it Chandra} MCD component
  (outer radius is set to infinity) at longer wavelengths.
Ionizing flux estimates for HeII, HeI and HI are shown by diamonds with error bars.
Asterisk corresponds to the optical source {\bf d} \citep{BFS} in
  the V band corrected for interstellar absorption $A_V = 1\magdot{.}54$.
MCD SEDs are shown (thin solid lines) for
  $\dot{m} = 0.01$ and masses 10, 100, 1000 and  $10^4~\Msun$.
The best-fit black body obtained during our {\it Cloudy} modelling (see
  section~\ref{sec:grid}) is shown by a dotted line.
{\it GALEX} photometric limits (FUV and NUV bands) are given for
  S/N=5, exposure time $10^5\,\rm s$.
Gray line shows the best-fit {\it Cloudy} model spectrum of the nebula.
}\label{fig:sedgalex}
\end{figure*}

Let us assume the SED of the central object smooth throughout the UV
range and interpolate betweeen the optical and EUV flux estimates with
a power-law (see figure~\ref{fig:sedgalex}). 
The predicted intrinsic (unabsorbed) flux
at 1000$\AAA$ is  $F_\lambda \sim 10^{-15} \ergf\AAA^{-1}$.
At shorter wavelengths radiation is highly absorbed by
neutral gas. 
Absorption by dust at $\lambda \sim 1000\div 2000\AAA$ may be estimated as
about $(3\div4)\times A_V \sim 5\magdot{\,}$ \citep{CCM}, so one
should expect at 1000$\AAA$ $F_\lambda \sim  10^{-17} \ergf\AAA^{-1}$,
that corresponds to $m_{AB} \sim 25\magdot{\,}$ (we use $m_{AB}$
definition from \cite{OG}).

Predicted UV flux is quite reachable for {\it GALEX} pointing
observations but too faint to be detected in the All-sky and Medium-sky imaging surveys
\citep{martin} mainly because of high interstellar absorption in UV
spectral range. Less absorbed ULXs might be quite achievable targets
if they are as bright as NGC6946~ULX-1 in the UV.
In figure~\ref{fig:sedgalex} we compare the $\nu F_\nu$ values with
 flux limits ($10^5\rm s$, $S/N = 5$) for {\it GALEX} photometry in FUV (1400-1700\AAA) and NUV
(2000-2700\AAA) bands. Signal-to-noise estimates were made with the
{\it GALEX} Exposure Time Calculator ({\it
  http://kaweah.caltech.edu:8000/ExpCalc.tcl}). 


Capabilities of the \hst\ instrumentation may be used as well to study
ULX sources, at least those less absorbed than NGC6946~ULX-1. 
\citet{liu_ngc5204} studied the UV counterpart of NGC5204~X-1 (U1
object) with STIS MAMA in the FUV
range and classified  the UV object as a B0Ia star with some
oddities like strong NV$\lambda$1240 emission that may be attributed to
an ambient HII region, irradiated accretion disc or corona. 
Some resonance lines like SiIV$\lambda$1400 and CIV$\lambda$1550 that
show P~Cygni-type profiles in  OB supergiant spectra are missing (or
present only as weak absorptions) in the spectra. 
\citet{liu_ngc5204} explain this effect by Roche lobe overflow by the
B supergiant donor star in a binary system.
This argumentation seems to be questionable because the winds of
early B-stars have terminal velocities about $1000$\kms, an
order of magnitude higher then virial velocities characteristic for HMXBs.
There is however an alternative explanation for the unusual absorption
spectrum of U1: P~Cygni profiles of UV lines formed in hot rarefied
atmospheres of OB~Ia/b supergiants transform into absorptions with
increasing mass loss \citep{hutchings}.
Mass loss rate needed for these effects to become important is of the
order $10^{-4}\Msun\rm yr^{-1}$. 
It is also possible that the OB-supergiant spectrum originates not
from the donor star but from the accretion disc wind atmosphere, that
is probable if the mass accretion rate is close to that observed in SS433.

Monochromatic flux at 2200\AAA reported by \citet{liu_ngc5204} for U1 is $F_\lambda \simeq 3.7
\times 10^{-17}$\ergf\AAA. 
The best-fit absorbing column measured for the X-ray source implies
$E(B-V)=0\magdot{.}28$. Galactic absorption is negligibly small
($E(B-V) =0\magdot{.}013$ according to \citet{schlegel_abs}).
Intrinsic UV luminosity of NGC5204~X-1 calculated using these values appears to
be $\nu L_\nu \sim 10^{38}$\ergl\ not taking into account absorption or $\sim
10^{39}$\ergl\ if reddening calculated from X-ray fitting is used.

Unfortunately, STIS stopped science operation in 2004 and its
potential for studying ULXs in the UV was lost.
ACS SBC and HRC cameras require exposures $\gtrsim 10^5\,\rm s$ to obtain
$S/N \gtrsim 1$ (exposure time calculator at {\it
  http://apt.stsci.edu/webetc/acs/acs\_img\_etc.jsp} was used) in the case of NGC6946~ULX-1
but may be used to study less absorbed bright UV counterparts.
Among the most perspective ULXs are NGC5204~X-1, NGC4559~X-7 \citep{ngc4559_soria}, HoII~X-1 \citep{lehmann}
and HoIX~X-1 \citep{hoix_miller} but we do not have any direct evidence
that these objects are as bright as NGC6946~ULX-1 (line luminosities
are generally lower).


The two most popular models of ULXs (IMBHs and supercritical accretion discs)
both predict strong UV/EUV emission. IMBHs are expected to
be bright UV sources due to the disc temperature dependence on the mass of
the accretor ($T_{in} \propto M^{-1/4}$). The greater is the mass of
the black hole the higher is the UV flux, approximately as $F
\propto M^{4/3}$ at a given
wavelength  for fixed dimensionless accretion rate
\citep{SS73,discbb}. Standard
accretion discs around IMBHs are expected to be truncated only at very
low temperatures, close to 1~eV (if their outer radii are due to tidal
truncation; see, for example, discussion in \citet{list}), and
must have therefore a distinguished spectral slope characteristic for
an MCD with an infinite outer radius, $F_\nu \propto \nu^{1/3}$
\citep{SS73}. 

Advective super-Eddington discs are often fitted by slim disc models
\citep{ACLS1988} predicting roughly flat SED in the X-ray band. 
On the other hand, supercritical accretion discs are expected to have
strong optically thick winds \citep{SS73} with photospheres having
blackbody-like spectrum peaking somewhere in the EUV/UV region
\citep{poutanen}. 
The exact structure of the accretion flow is
irrelevant for the EUV/UV SED because most of the EUV radiation comes from the
wind photosphere. In the {\it GALEX} range the spectrum is expected to
have a Rayleigh-Jeans shape $F_\nu \propto \nu^2$. 


We conclude that detecting ULXs in the {\it GALEX} UV
range and measuring the spectral slope may help to distinguish between these two models
of ULXs (IMBH or supercritical accretion). 
At least for some subsample of ULXs (those having indications
for a hard photoionizing continuum) different models predict different
spectral slopes in the UV range (see figure~\ref{fig:sedgalex}).

The nebula may contribute to the UV emission from the source.
Cloudy model predicts total flux in the GALEX range about an order of
magnitude less than that expected from the ULX itself.
Best-fit model predicts $L(Ly\alpha) \simeq 26 L(H\beta) \simeq
2\times 10^{39}\ergl$ but most of the L$\alpha$-quanta are likely to
be absorbed by neutral interstellar medium. 
Expected contribution from the nebula is shown in figure \ref{fig:sedgalex}.
Higher fluxes may appear if some additional amount of gas with at
least comparable emission measure and much higher temperature is
present (much
cooler gas does not contribute to the continuum in the UV range). 
But the optical spectra contain no signatures of high-temperature
plasma such as coronal lines of very high-ionization element species.
Therefore nebular emission is unlikely to give a significant
contribution to the {\it GALEX} UV in the particular case of MF16.

\subsection{MF16 as a Jet-Blown Shell}\label{sec:shell}

From the high spatial resolution {\it HST} images \citep{BFS} it can
be seen that the nebula is elongated with an axis ratio no less than
$\sim 1.5$. Either an underlying density gradient or highly anisotropic energy injection in the ISM is needed
to explain the observed morphology.

If a shell evolves in a strong density gradient, one side expands with
a much higher velocity making the nebula much more extended
 in one direction then in the other. 
Central position of the optical star {\bf d} is difficult to explain
in this scope. Density gradient probably plays significant
role only in making one side of MF16 much brighter. 
This is expected \citep{maciej} even for very shallow density
gradients. 

In all the self-similar solutions usually applied to expanding shells
\citep{Lozinsk} and
wind-blown bubbles \citep{castor}, energy and mechanical luminosity appear with
very low exponents. In particular, for the Sedov solution for an
adiabatic-stage SNR the radius of the shell depends as $R \propto
E^{1/5}$ on the energy input. 
Therefore to achieve 50\% variation in the radius of the nebula,  
energy injected per unit solid angle (or the initial ISM density) must
change no less than by an order of magnitude. 
However, if the central machine is a source
of well-collimated wind or jets, the shape is naturally explained.

Jet activity is unlikely to be present in standard discs (expected in
the framework of IMBH model) and may be
considered an argument for supercritical accretion. 
We do see a pair of powerful jets and an elongated nebula W50 in the case of
SS433.
However, if NGC6946~ULX-1 is a supercritical accretor viewed
face-on (at inclination $i \lesssim 20^\circ$)
the direction of jets must be close to the line of sight.
In that case the actual length of MF16 should be
several times (about $\sin^{-1} i$) greater, approaching the size of W50. 

NGC6946 is a seen nearly face-on:  according to \citet{gordon68}, its inclination is in the range $16\div
27^\circ$.  
If the actual size of MF16 in the direction perpendicular to the disc plane
of the host galaxy is of the order 100\pc\ as
in the case of W50 then the nebula is likely to evolve in an ISM with a
considerable density gradient.
The thickness of the gaseous disc in NGC6946 probably does not differ too much
from the Galactic neutral hydrogen disc thickness $ \simeq 300\,\pc$ \citep{lockman}.
The expected effects are similar to those observed in the case of W50
that evolves in a strong ISM density gradient \citep{lockman_w50}.
It is quite probable that the asymmetry of MF16 seen in {\it HST}
images
originates from the
interaction of one of the jets with denser material closer to the
galactic plane.




\section{Conclusions}

In our optical spectra of MF16 we detect more than 30 
emission lines including several high-excitation lines of \heii\ and
\ariv\ and a rich spectrum of moderately high excitation \feiii\
lines. \heii\ lines are narrow (broadened by $\lesssim 300\kms$). We
also do not detect any Wolf-Rayet features. Therefore we
suggest that the \heii$\lambda$4686 line is of nebular origin. 

Large intensities of iron lines indicate moderately low depletion of the
element into dust ($30\div 50\%$ in the gas phase), probably due to dust destruction in
shock waves. Destruction is probably incomplete because {\it Cloudy}
models overestimate the intensities of \feiii\ lines.

We find the electron density in the nebula $n_e = 570\, \pm 60\,\cmc$
(\sii$\lambda$6717,6731), electron temperatures for different ions are
$T(\oiii) = 17\,700\pm1\, 200\,\rm K$, 
 $T(\nii) = 15\,600\pm 2\,000\,\rm K$ 
 and $T(\sii) = 9\,000\pm 1000\,\rm K$ 
indicating the presence of regions with
different electron temperatures. Total hydrogen emitting gas mass is $M \sim 900\,\Msun$. 
Interstellar absorption is $A_V \simeq 1\magdot{.}34$
in traditional H$\alpha$/H$\beta$=3 assumption. 
More realistic value  H$\alpha$/H$\beta$=2.8
results in a higher extinction value $A_V = 1\magdot{.}54$.

The observed line luminosities and diagnostic line ratios appear
to be inconsistent with excitation and ionization by shock waves,
therefore we suggest an EUV source responsible for 
powering the nebula. Photoionization modelling with {\it Cloudy} as well as Zanstra
estimates suggest the central source must be 
ultraluminous not only in X-rays but also in the UV/EUV range
emitting about $10^{40}$\ergl\ in the spectral range $100-1000\AAA$.
Using the observed X-ray spectrum,
ionizing flux estimates from \heii, \hei\ and H lines and the optical
point-like counterpart (star {\bf d}) we reconstruct
the SED of the central object from X-rays to the optical.
The derived spectrum is roughly flat, $\nu L_{\nu} \sim const$. 

Both most popular models of ULXs (IMBHs and supercritical accretion discs) predict high UV/EUV luminosities.
However these two models predict different spectral slopes in the UV region. 
We conclude that measuring the spectral slope will help to distinguish between the two
models as well as to determine the parameters of the successfull
model such as the black hole mass in the case of IMBH or accretion 
rate in the case of a supercritical accretion.

Actual excitation and ionization conditions in MF16 may be much more
complicated. Though introducing a bright EUV source is a possible way
to explain the spectrum, we suggest that other explanations such as
high pre-shock density and multiple shocks with different velocities
are not completely excluded. Observations with higher angular and
spectral resolution (but also higher S/N ratio than that used
by \citet{dunne}) may provide additional information about the
enigmatic nebula MF16.

\bigskip

We thank V. Afanasiev and N. Borisov for assistance with the
 observations and the anonymous referee for valuable comments and
 suggestions. 
 This work was supported by
 the Russian RFBR grants 06-02-16865 and 07-02-00909 and the RFBR/JSPS grant
 05-02-19710.
TK is supported by a 21st Century COE Program at Tokyo Tech
``Nanometer-Scale Quantum Physics'' by the Ministry of Education,
Culture, Sports, Science and Technology.  This work is supported by the
Japan-Russia Research Cooperative Program of Japan Society for the
Promotion of Science.

\begin{table*}
\caption{Observational logs}\label{tab:obstab}
\center{
         \begin{tabular}{lcr}
               Spectrograph & MPFS & SCORPIO (long-slit mode)\\
               Date & 2005/17/01 & 2005/10/06\\
               Total exposure (s) & 5829 & 5400 \\
               Spectral range ($\AAA$) & 4000-7000 & 3900-5700\\
               Spectral resolution ($\AAA$) & 6 & 5 \\
	       Seeing, arcsec &      1.7 & 1.6\\
            \noalign{\smallskip}
         \end{tabular}
}
\end{table*}

\begin{table*}
\caption{MF16 emission line parameters.
Line intensities in H$\beta$ units (the uncertainties do 
not account for the the H$\beta$ flux uncertainty), radial velocities
and FWHMs.   
\label{tab:lines}
}
\small
\center{ 
\begin{tabular}{lcccc}
\hline
\noalign{\smallskip}
line & $\frac{F(\lambda)}{F(H_\beta)}$ & $\frac{F(\lambda)}{F(H_\beta)}$ & $V,km\,s^{-1}$ 
& $FWHM,\rm\AA$ \\
& & (unreddened) & & \\ 
\hline
\noalign{\smallskip}
\hline
\noalign{\smallskip}
$[SII]\lambda 4068$   &0.200$\pm$0.013    &0.276$\pm$0.018
&-100$\pm$20&8.3$\pm$0.7 \\ 
$H\delta$             &0.228$\pm$0.013    &0.311$\pm$0.018   &-66$\pm$11&6.6$\pm$0.4 \\ 
$[TiIII]\lambda 4161$?&0.018$\pm$0.008    &0.024$\pm$0.010   &0$\pm$40&3.8$\pm$1.2 \\ 
$SIII\lambda4284$?    &0.074$\pm$0.012    &0.094$\pm$0.015   &-50$\pm$40&11.2$\pm$1.3 \\ 
$H\gamma$             &0.408$\pm$0.013    &0.510$\pm$0.016   &-47$\pm$14&6.2 $\pm$0.2 \\ 
$[OIII]\lambda 4363$  &0.146$\pm$0.016    &0.181$\pm$0.019   &-73$\pm$14&7.1$\pm$0.5 \\ 
$[FeII]\lambda 4414$? &0.039$\pm$0.010    &0.047$\pm$0.012   &-10$\pm$50&8.6$\pm$1.6 \\ 
$HeI\lambda 4471$     &0.038$\pm$0.006    &0.045$\pm$0.007   &-56$\pm$20&6.0$\pm$0.7 \\ 
$[FeIII]\lambda 4658$ &0.066$\pm$0.011    &0.072$\pm$0.012   &10$\pm$30&9.3$\pm$1.2 \\ 
$HeII\lambda 4686$    &0.205$\pm$0.009    &0.222$\pm$0.010     &-5$\pm$18&6.2$\pm$0.2 \\ 
$[FeIII]\lambda 4701$ &0.017$\pm$0.008    &0.018$\pm$0.008   &-35$\pm$50&5.1$\pm$1.7 \\ 
$HeI\lambda 4713$ (+$[ArIV]\lambda4711$?) &0.021$\pm$0.008   &0.021$\pm$0.008&-70$\pm$40&5.3$\pm$1.6 \\ 
$[ArIV]\lambda 4740$  &0.009$\pm$0.004    &0.009$\pm$0.004   &-50$\pm$30&3$\pm$1 \\ 
$H\beta$              &1.000$\pm$0.017    &1.000$\pm$0.017   &-28$\pm$4&6.2$\pm$0.2 \\ 
$[FeIII]\lambda 4881$ &0.018$\pm$0.008    &0.018$\pm$0.008   &-86$\pm$20&4.8$\pm$1.0 \\ 
$HeI\lambda 4922$     &0.011$\pm$0.005    &0.011$\pm$0.005   &-10$\pm$30&3.9$\pm$1.2 \\ 
$[FeIII]\lambda 4936$ &0.008$\pm$0.005    &0.008$\pm$0.005   &-60$\pm$50&4.0$\pm$1.5 \\ 
$[OIII]\lambda 4959$  &2.337$\pm$0.050    &2.24$\pm$0.05     &-23$\pm$3&6.2$\pm$0.1 \\ 
$[FeIII]\lambda 4986$ &0.044$\pm$0.010    &0.042$\pm$0.009   &50$\pm$100&4.0$\pm$1.5 \\ 
$[OIII]\lambda 5007$  &6.94$\pm$0.14      &6.50$\pm$0.13     &-17.4$\pm$2.5&6.2$\pm$0.1 \\ 
$[FeIII]\lambda 5033$?&0.013$\pm$0.005    &0.012$\pm$0.005   &50$\pm$30&4.1$\pm$1.4 \\ 
$[FeII]\lambda 5158$? &0.073$\pm$0.005    &0.064$\pm$0.004   &11$\pm$9&7.8$\pm$0.4 \\ 
$[NI]\lambda 5199$    &0.152$\pm$0.005    &0.131$\pm$0.004   &-34$\pm$5 &7.1$\pm$2.0 \\ 
$[FeII]\lambda 5262$? &0.072$\pm$0.014    &0.061$\pm$0.012   &140$\pm$90&18$\pm$2 \\ 
$[FeII]~\!\!+~\!\!\!FeIII\lambda5270$
                      &0.018$\pm$0.008    &0.015$\pm$0.007   &60$\pm$30 &5.1$\pm$1.7 \\ 
$FeIII\lambda 5300$?  &0.011$\pm$0.005    &0.009$\pm$0.004   &100$\pm$50 &5$\pm$2 \\ 
$HeII\lambda 5411$    &0.021$\pm$0.005    &0.017$\pm$0.004   &-10$\pm$20&5.6$\pm$1.0 \\ 
$[NII]\lambda 5755$   &0.118$\pm$0.015    &0.084$\pm$0.011   &-70$\pm$30&16.0$\pm$1.5 \\ 
$HeI\lambda 5876$     &0.114$\pm$0.024    &0.078$\pm$0.016   &-9$\pm$20&7.0$\pm$1.1  \\ 
$[OI]\lambda 6300$    &1.421$\pm$0.083    &0.89$\pm$0.05     &25$\pm$9&10.1$\pm$0.4  \\ 
$[OI]\lambda 6364$    &0.480$\pm$0.029    &0.296$\pm$0.018   &12$\pm$9&9.7$\pm$0.4  \\ 
$[NII]\lambda 6548$   &1.385$\pm$0.027    &0.822$\pm$0.016   &-18.7$\pm$1.3&7.6$\pm$0.1  \\ 
$H\alpha$             &4.728$\pm$0.087    &2.80$\pm$0.05     &-15.7$\pm$2.1&7.8$\pm$0.1  \\ 
$[NII]\lambda 6583$   &4.156$\pm$0.082    &2.45$\pm$0.05     &-18.6$\pm$2.2&7.6$\pm$0.1  \\ 
$HeI\lambda 6678$     &0.060$\pm$0.032    &0.035$\pm$0.018   &-18$\pm$9&14.0$\pm$5.2\\ 
$[SII]\lambda 6717$   &2.457$\pm$0.024    &1.412$\pm$0.014   &-37.5$\pm$1.1&7.7$\pm$0.1  \\ 
$[SII]\lambda 6731$   &2.348$\pm$0.025    &1.346$\pm$0.014   &-32.8$\pm$1.2&8.2$\pm$0.1  \\ 
\hline
\noalign{\smallskip}
\end{tabular}
}
\normalsize
\end{table*}

\begin{table*}
\caption{Fitting emission line fluxes with {\it Cloudy} models. Emission
  line fluxes are
  given in $10^{-16}\ergf$.
\label{tab:linefit}
}
\center{
         \begin{tabular}{l c ccc c ccc} 
line &&  $F_{obs}$ & $F_{model}$ & $\Delta\chi^2$ &&   $F_{obs}$ &
	   $F_{model}$ & $\Delta\chi^2$ \\
$A_V,\magdot{\,}$ && \multicolumn{3}{c }{$ 1.34$} &&  \multicolumn{3}{c}{$1.54$}\\
\feiii$\lambda$4658  && 11$\pm$2  &
	   30 & 90 &&  18$\pm$3 & 35 & 32 \\ 
\heii$\lambda$4686  && 44$\pm$2  & 36
	   & 16 &&  56$\pm$2 & 44 & 36 \\ 
H$\beta$  && 200$\pm$3  & 205  & 2  &&  249$\pm$4 & 246 & 0.56 \\ 
\oiii$\lambda$5007  && 1310$\pm$30  &
	   1089 & 54  &&  1620$\pm$30 & 1520 & 11 \\ 
\hei$\lambda$5876  && 16$\pm$3  & 27
	   & 13 &&  20$\pm$4 & 32 & 9 \\ 
\nii$\lambda$6584  && 526$\pm$10  &
	   484 & 15 &&  611$\pm$12 & 527 & 49 \\ 
\sii$\lambda$6717  && 304$\pm$3  &
	   334 & 100 &&  352$\pm$3 & 367 & 25 \\ 
\sii$\lambda$6731  && 290$\pm$3  &
	   316 & 75 &&  335$\pm$4 & 348 & 11 \\ 
$F,\ergf$ && \multicolumn{3}{c }{$0.55^{+0.1}_{-0.05}$} &&  \multicolumn{3}{c}{$0.6^{+0.1}_{-0.1}$}\\
$T,10^5\,K$ && \multicolumn{3}{c}{$5.05^{+0.15}_{-0.05}$} &&
	   \multicolumn{3}{ c}{$5.1^{+0.15}_{-0.05}$}\\
$\chi^2 / DOF$ &&  \multicolumn{3}{c }{62}  &&  \multicolumn{3}{c}{25} \\
	 \end{tabular}
}
\end{table*}

\end{document}